# On the System-level Performance of Coordinated Multi-point Transmission Schemes in 5G NR Deployment Scenarios

*Siva Muruganathan, Sebastian Faxér, Simon Järmyr, Shiwei Gao, and Mattias Frenne*

*Ericsson*

*Email: {siva.muruganathan, sebastian.faxer, simon.jarmyr, shiwei.gao, mattias.frenne}@ericsson.com*

*Abstract*—This paper investigates the system-level performance of dynamic point selection (DPS) and non-coherent joint transmission (NC-JT) coordinated multi-point transmission (CoMP) schemes under different 5G NR deployment scenarios using state-of-the-art system-level simulations. It is observed that at a mid-band carrier frequency, NC-JT does not provide performance gains over DPS or single transmission/reception point (TRP) transmission unless the channel from a TRP is rank deficient. Therefore, benefits with NC-JT are more likely to be found in indoor deployment scenarios where the TRPs are typically equipped with only 2 transmit antenna ports, whereas benefits are less likely to be observed in macro-cell deployments where TRPs typically have a larger number of antennas ports. It is further observed that NC-JT gains tend to diminish with increasing system load, partly due to increased interference, which typically lowers the transmission rank.

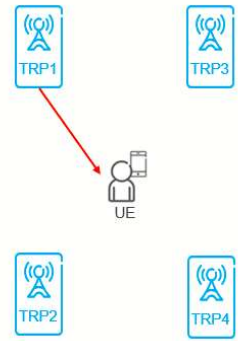

Figure 1: Illustration of single-TRP transmission

## I. INTRODUCTION

In recent years, the demand for mobile data traffic has grown at an unprecedented and accelerating rate, approaching a year-on-year growth in data traffic of almost 100% [1]. This increasing demand for traffic has been one of the key drivers for the development of fifth-generation (5G) cellular networks and in particular the 3GPP 5G New Radio (NR) standard, with its first release (Rel-15) completed in June 2018 [2]. Compared to the 4G LTE-A standard, 5G NR brings improvements in user experienced data rate, network capacity, latency, energy efficiency and reliability [3]. One principal component of the 5G NR standard is the use of multi-input-multi-output (MIMO) transmission techniques for improved coverage and spectral efficiency, both in traditional sub-6 GHz and in millimeter wave (mmWave) deployments.

The 5G NR standard will continue to evolve and improve. In the ongoing 3GPP Rel-16 work, enhancements to NR MIMO functionality is considered [4]. One candidate scheme for improving spectral efficiently is the use of coordinated multi-point (CoMP) transmission. In traditional cellular networks, the user equipment (UE) is only connected to a single transmission/reception point (TRP) of the network at a time and each TRP makes independent scheduling, precoding and resource allocation decision, as is illustrated in Figure 1. For CoMP, on the other hand, multiple TRPs cooperate and coordinate their transmissions such that a UE can receive transmissions from multiple TRPs simultaneously.

Several different CoMP schemes are being considered, each with different characteristics in terms of achievable performance and demand for coordination. The baseline Rel-15 NR scheme uses dynamic point selection (DPS), where a UE can be dynamically scheduled to be served by one of multiple TRPs in a coordination cluster, as illustrated in Figure 2a. Such dynamic scheduling can be on a per transmission time interval (TTI) basis. This enables the network to serve the UE by the TRP which momentarily offers the best channel conditions to the UE, or to perform dynamic load balancing between the TRPs. Support for DPS was enabled already in LTE Rel-11 with the introduction of the quasi-co-location (QCL) property between reference signals and multiple channel state information (CSI) processes [5].

Another related CoMP feature is dynamic point blanking (DPB), where a joint multi-TRP scheduler can make a dynamic decision to not schedule any UEs from one or more TRPs, thereby reducing interference to the UEs served by the remaining TRPs, as is illustrated in Figure 2b. Typically, DPS and DPB are operated simultaneously and correspond to different scheduling hypotheses of a single joint scheduler.

While for DPS/DPB, the UE is only receiving transmissions from a single TRP at a given time, in joint transmission (JT), the UE can receive transmissions from multiple TRPs simultaneously. JT can further be broken down into two flavors, non-coherent JT (NC-JT) and coherent JT (C-JT). For C-JT, the same precoding layer(s) are transmitted from the multiple cooperating TRPs with the intention for the transmissions to coherently add up at the receiving UE, as is illustrated in Figure 2d. In order to accomplish this, tight synchronization is required between the TRPs and a high CSI accuracy is needed in order to design the precoding weights, which makes C-JT difficult to practically implement and requires large CSI feedback overhead.

In NC-JT, on the other hand, the requirement for synchronization and CSI accuracy is comparably lower, since each TRP is transmitting different layers, as is illustrated in Figure 2c. Limited support for NC-JT was introduced in the feCoMP work item (WI) for LTE Rel-15 [6] and is further considered also for NR in the Rel-16 eMIMO WI. Currently, there are no plans in 3GPP to introduce support for C-JT for LTE or NR.

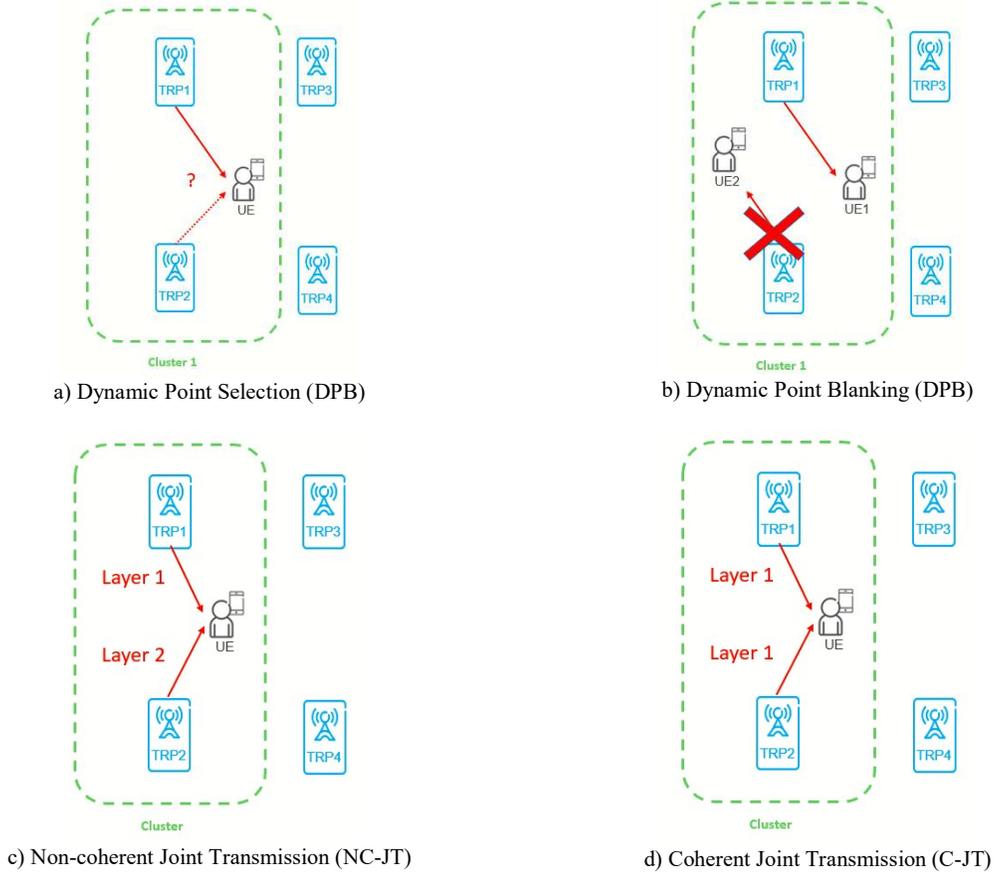

Figure 2: Illustration of four CoMP transmission schemes

a) Dynamic Point Selection (DPB)

b) Dynamic Point Blanking (DPB)

c) Non-coherent Joint Transmission (NC-JT)

d) Coherent Joint Transmission (C-JT)

In this paper, the performance of DPS and NC-JT CoMP schemes are evaluated and analyzed in a 5G NR context with realistic system-level simulations using state-of-the-art 5G channel models and deployment scenarios. The remainder of the paper is organized as follows: Section II gives a description of the investigated scenarios, Section III presents evaluation results, and Section IV provides concluding remarks.

## II. 5G NR Deployment Scenarios

This section describes four different 5G NR deployment scenarios with different characteristics for which evaluation results are presented in Section III.

### A. Indoor Hotspot Scenario at 4 GHz Carrier Frequency

First, consider the indoor hotspot (InH) scenario at 4 GHz carrier frequency as defined in 3GPP technical report TR38.802, section A.2 [7]. This scenario consists of a deployment with 12 ceiling mounted TRPs that are placed in a rectangular area of 120m × 50m.

Each TRP is assumed to be equipped with $N_{Tx}$ transmit antenna ports that are facing down from the ceiling. For the evaluation results in Section III, results for $N_{Tx} = 2$ and $N_{Tx} = 4$ transmit antenna ports per TRP are provided. Note that these numbers of antenna ports are typical for InH TRPs in mid-band carrier frequencies (for example, 4 GHz) due to their small form factor. At the UE side, each UE is assumed to be equipped with 4 receive antennas and an MMSE-IRC receiver. Note that with 4 receive antenna ports, the UE is capable of receiving up to 4 MIMO layers, i.e., a maximal rank of 4. With $N_{Tx} = 2$, a TRP can only transmit up to 2 MIMO layers or maximal rank 2, thus there is a rank gap between what the UE is capable of receiving and what a single TRP can transmit. This case is referred to as "rank deficient" induced by available number of transmit antennas. Note that the channel can also be rank deficient due to the property of the channel, in, e.g., line-of-sight conditions.

For evaluations involving DPS and NC-JT, the TRPs are partitioned into different coordination clusters wherein two adjacent TRPs form one coordination cluster. It is assumed that each UE is associated with one coordinated cluster. The TRPs belonging to one coordination cluster are connected via ideal backhaul (i.e., with zero latency). This enables coordinated scheduling with a single scheduler per coordination cluster. A UE can hence be scheduled from either one of the TRPs (using DPS) or both TRPs (using NC-JT) in the coordination cluster based on, e.g., proportional fairness (PF) scheduling metrics.

## B. Dense Urban Macro Scenario at 4 GHz Carrier Frequency

Consider next the dense urban (DU) macro scenario at 4 GHz carrier frequency as defined in 3GPP technical report TR38.802, section A.2 [7] although here only the macro layer is modelled. In our evaluations, 19 macro sites were assumed with each site consisting of 3 macro sectors (i.e., macro TRPs).

For the DU scenario, two different numbers of transmit antenna ports were considered: $N_{Tx}$ = 2, and $N_{Tx}$ = 4. At the UE side, each UE is assumed to be equipped with 4 receive antennas and an MMSE-IRC receiver.

When evaluating the DPS and NC-JT schemes, it was assumed that the macro TRPs belonging to the same macro site are connected via ideal backhaul and comprise one coordination cluster, and that each UE is associated with one coordination cluster. Due to the ideal backhaul assumption, it is possible to use a single scheduler per coordination cluster, where the scheduler is responsible scheduling data to the UE using either DPS or NC-JT.

## C. Indoor Hotspot Scenario at 30 GHz Carrier Frequency

The third scenario considered in our evaluations is the InH scenario at 30 GHz carrier frequency where analog, or time domain, beamforming is assumed at the transmitter and the receiver. The placement of TRPs for this scenario is like that described in Section IIA.

Each TRP is assumed to be equipped with a 4x4 cross-polarized antenna array capable of generating 16 analog beams for transmission. It is assumed that $N_{Tx}$ = 2 transmit antenna ports are used per analog beam. The UEs are assumed to be equipped with two panels facing in opposite directions. Each of the panels at the UE is assumed to contain a 2x4 cross polarized antenna array with 8 analog beams. Each of the analog beams associated with one panel has 2 receive antenna ports. Hence, there are 4 receive antenna ports in total across two analog beams received using the two panels. In addition, the UE is assumed to be equipped with a MMSE-IRC receiver.

The coordination cluster setup is similar to Section IIA where two adjacent TRPs form one coordination cluster.

## D. Dense Urban Macro Scenario at 30 GHz Carrier Frequency

The last scenario considered in our evaluations is the DU scenario at 30 GHz carrier frequency where analog beamforming is assumed at the transmitter and the receiver. The placement of TRPs for this scenario is like that described in Section IIB, with the exception that 7 sites are considered instead of 19.

In this scenario, each TRP is assumed to be equipped with an 8x8 cross-polarized antenna array with 32 analog beams available for transmission. Each analog beam is assumed to contain $N_{Tx}$ = 2 transmit antenna ports. The UEs are assumed to be equipped with two panels facing in opposite directions with a setup like that described in Section IIC.

The coordination cluster setup is like Section IIB where three macro TRPs belonging to the same macro site form one coordination cluster.

## III. SYSTEM LEVEL EVALUATON RESULTS

Evaluation results for the scenarios described in Section II are provided in this section. These results are generated using a system-level simulator where the traffic is generated using 3GPP FTP Model 1 [8] with a packet size of 0.5 Megabytes.

The performance is measured in terms of mean and cell edge user-perceived throughput (UPT), where cell edge UPT is defined as the 5$^{th}$ percentile UPT, and is presented as relative gains in mean and cell edge UPT with respect to a single TRP baseline (i.e., no CoMP). The performance is evaluated for different system loads in the baseline system. The system loads are characterized by average percentage of system resource utilization (RU) across all TRPs. In these evaluations, 10%, 20%, and 40% system RU of the baseline scheme are considered.

### A. Indoor Hotspot Results at 4 GHz Carrier Frequency

The results for InH at 4 GHz carrier frequency are given in Figure 3 and Figure 4. From Figure 3, it is observed that with $N_{Tx}$ = 2 transmit antenna ports per TRP, NC-JT can achieve notable mean throughputs gains over both single TRP transmission and DPS schemes at 10% RU. The mean throughput gain of NC-JT is moderate at 20% RU, and beyond 20% RU NC-JT performs poorly when compared to both single-TRP transmission and DPS.

In contrast to the results in Figure 3, the results from Figure 4 with $N_{Tx}$ = 4 transmit antenna ports per TRP indicate that NC-JT performs poorly when compared to both DPS and single-TRP transmission for all RU values.

The main reason for poor performance of NC-JT is likely that with $N_{Tx}$ = 4 transmit antenna ports per TRP, a maximum rank (or number of MIMO layers) of 4 can already be achieved from a single TRP and using NC-JT does not provide any benefits. Conversely, using NC-JT results in data being transmitted from two TRPs which increases the interference in the system which leads to poor performance of NC-JT. With $N_{Tx}$ = 2 transmit antenna ports per TRP, however, the transmission from a single TRP is rank deficient and is limited to a maximum rank of 2. In this case, transmission from two TRPs using NC-JT can increase the maximum rank to 4 which can be beneficial at low loads where the interference tends to be low.

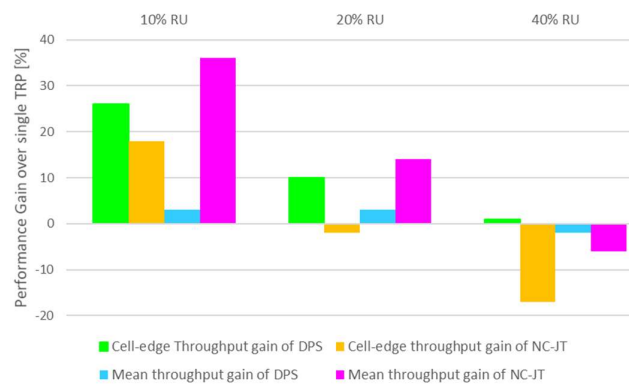

Figure 3. Results for InH at 4 GHz carrier frequency with $N_{Tx}$ = 2 transmit antenna ports per TRP and 4 receive antenna ports at UE.

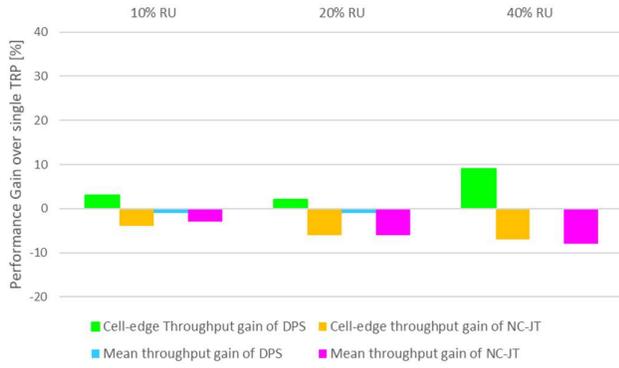

Figure 4. Results for InH at 4 GHz carrier frequency with $N_{Tx} = 4$ transmit antenna ports per TRP and 4 receive antenna ports at UE.

### B. Dense Urban Results at 4 GHz Carrier Frequency

The results for DU at 4 GHz carrier frequency are given in Figure 5 and Figure 6.

Similar to the observations in Section IIIA, with $N_{Tx} = 2$ transmit antenna ports per TRP in the DU scenario, NC-JT has some mean throughput gain over single TRP transmission and DPS at 10% RU. However, above 20% RU, NC-JT performs poorly compared to both single TRP transmission and DPS.

With $N_{Tx} = 4$ transmit antenna ports per TRP in the DU scenario, NC-JT does not provide any mean throughput gains over DPS and single TRP transmission. This result is also consistent with the observation in Section IIIA.

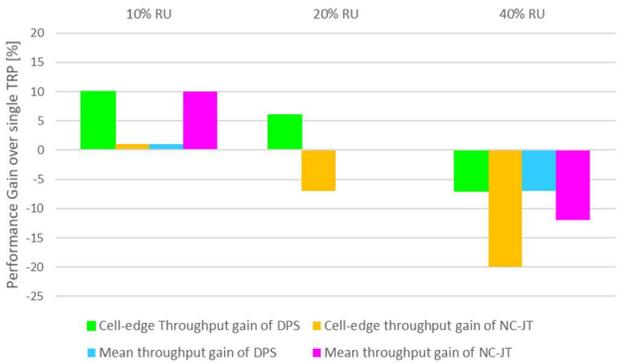

Figure 5. Results for DU at 4 GHz carrier frequency with $N_{Tx} = 2$ transmit antenna ports per TRP and 4 receive antenna ports at UE.

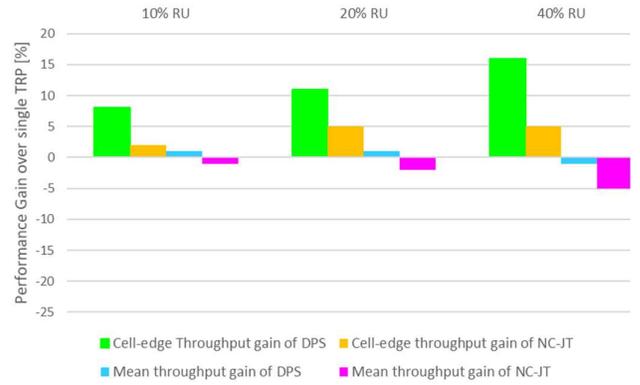

Figure 6. Results for DU at 4 GHz carrier frequency with $N_{Tx} = 4$ transmit antenna ports per TRP and 4 receive antenna ports at UE.

### C. Indoor Hotspot Results at 30 GHz Carrier Frequency

The results for InH at 30 GHz carrier frequency are given in Figure 7. From these results, it is noted that NC-JT provides similar cell-edge performance to that of DPS. However, NC-JT provides a fair mean throughput gain over DPS. The reason is that DPS is limited to a maximum rank of 2 since only 2 transmit antenna ports are possible per analog beam transmitted from a single TRP. On the other hand, by transmitting from two TRPs, NC-JT is able to deliver up to a maximum rank of 4 to a UE which results in the mean throughput gain of NC-JT over DPS. It should however be emphasized that NC-JT gains tend to diminish with increasing RU. This is because a higher RU leads to more interference in the system which typically lowers the desired transmission rank, and hence also lowers the potential benefit of NC-JT.

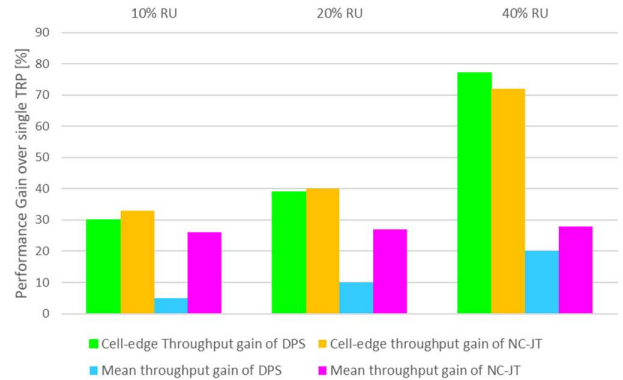

Figure 7. Results for InH at 30 GHz carrier frequency.

## D. Dense Urban Results at 30 GHz Carrier Frequency

The results for DU at 30 GHz carrier frequency are given in Figure 8. Similar to the results in Section IIIC, it is observed that NC-JT provides a fair mean throughput gain over DPS. This again is due to NC-JT being able to deliver up to a maximum rank of 4 to a UE while DPS is limited to a maximum rank of 2.

## IV. CONCLUSIONS

In this paper, the performance of NC-JT under different 5G NR deployment scenarios were evaluated. From the evaluation results, the following are observed:

At a mid-band carrier frequency of 4 GHz, NC-JT does not provide performance gains over DPS and single-TRP transmission when the single TRP channel is not rank deficient, i.e., each TRP can provide the maximum rank that a UE can receive. However, if each TRP is equipped with a smaller number of transmit antenna ports than the number receive antennas ports available at the UE, or if line-of-sight probability is high, some benefits may be possible for NC-JT over DPS and single-TRP transmission.

At a mid-band carrier frequency of 4 GHz, most of the NC-JT benefit for NC-JT are found in indoor scenarios where the TRPs are most likely to be equipped with 2 transmit antenna ports while the UEs are equipped with 4 receive antenna ports.

At a high-band carrier frequency of 30 GHz, NC-JT outperforms both DPS and single-TRP transmission as both of these schemes are limited to a maximum transmission rank of 2. On the other hand, by transmitting from two TRPs, NC-JT is able of delivering up to a maximum rank of 4.

NC-JT gains tend to diminish with increasing system load as higher system load leads to higher interference which typically lowers the desired transmission rank.

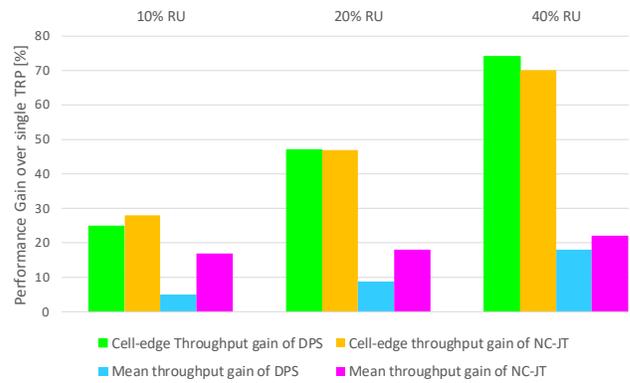

Figure 8. Results for DU at 30 GHz carrier frequency.